\documentclass[reprint, amsmath, amssymb, pra, superscriptaddress]{revtex4-2}

\usepackage{graphicx}
\usepackage{bm}
\usepackage{color}
\usepackage{siunitx}
\usepackage[colorlinks,linkcolor=blue,anchorcolor=blue,citecolor=blue,urlcolor=blue]{hyperref}

\newcommand{\supplement}{Appendix}
\newcommand{\tr}{\mathop\text{tr}\nolimits}

\begin{document}

\title{Probing two driven double quantum dots strongly coupled to a cavity}

\author{Si-Si Gu}
\affiliation{CAS Key Laboratory of Quantum Information, University of Science and Technology of China, Hefei, Anhui 230026, China}
\affiliation{CAS Center for Excellence in Quantum Information and Quantum Physics, University of Science and Technology of China, Hefei, Anhui 230026, China}

\author{Sigmund Kohler}
\email[]{sigmund.kohler@csic.es}
\affiliation{Instituto de Ciencia de Materiales de Madrid, CSIC, E-28049 Madrid, Spain}

\author{Yong-Qiang Xu}
\affiliation{CAS Key Laboratory of Quantum Information, University of Science and Technology of China, Hefei, Anhui 230026, China}
\affiliation{CAS Center for Excellence in Quantum Information and Quantum Physics, University of Science and Technology of China, Hefei, Anhui 230026, China}

\author{Rui Wu}
\affiliation{CAS Key Laboratory of Quantum Information, University of Science and Technology of China, Hefei, Anhui 230026, China}
\affiliation{CAS Center for Excellence in Quantum Information and Quantum Physics, University of Science and Technology of China, Hefei, Anhui 230026, China}

\author{Shun-Li Jiang}
\affiliation{CAS Key Laboratory of Quantum Information, University of Science and Technology of China, Hefei, Anhui 230026, China}
\affiliation{CAS Center for Excellence in Quantum Information and Quantum Physics, University of Science and Technology of China, Hefei, Anhui 230026, China}

\author{Shu-Kun Ye}
\affiliation{CAS Key Laboratory of Quantum Information, University of Science and Technology of China, Hefei, Anhui 230026, China}
\affiliation{CAS Center for Excellence in Quantum Information and Quantum Physics, University of Science and Technology of China, Hefei, Anhui 230026, China}

\author{Ting Lin}
\affiliation{CAS Key Laboratory of Quantum Information, University of Science and Technology of China, Hefei, Anhui 230026, China}
\affiliation{CAS Center for Excellence in Quantum Information and Quantum Physics, University of Science and Technology of China, Hefei, Anhui 230026, China}

\author{Bao-Chuan Wang}
\affiliation{CAS Key Laboratory of Quantum Information, University of Science and Technology of China, Hefei, Anhui 230026, China}
\affiliation{CAS Center for Excellence in Quantum Information and Quantum Physics, University of Science and Technology of China, Hefei, Anhui 230026, China}

\author{Hai-Ou Li}
\affiliation{CAS Key Laboratory of Quantum Information, University of Science and Technology of China, Hefei, Anhui 230026, China}
\affiliation{CAS Center for Excellence in Quantum Information and Quantum Physics, University of Science and Technology of China, Hefei, Anhui 230026, China}

\author{Gang Cao}
\email[]{gcao@ustc.edu.cn}
\affiliation{CAS Key Laboratory of Quantum Information, University of Science and Technology of China, Hefei, Anhui 230026, China}
\affiliation{CAS Center for Excellence in Quantum Information and Quantum Physics, University of Science and Technology of China, Hefei, Anhui 230026, China}

\author{Guo-Ping Guo}
\email[]{gpguo@ustc.edu.cn}
\affiliation{CAS Key Laboratory of Quantum Information, University of Science and Technology of China, Hefei, Anhui 230026, China}
\affiliation{CAS Center for Excellence in Quantum Information and Quantum Physics, University of Science and Technology of China, Hefei, Anhui 230026, China}
\affiliation{Origin Quantum Computing Company Limited, Hefei, Anhui 230088, China}

\date{\today}

\begin{abstract}
We experimentally and theoretically study a driven hybrid circuit quantum
electrodynamics (cQED) system beyond the dispersive coupling regime.
Treating the cavity as part of the driven system, we develop a theory
applicable to such strongly coupled and to multi-qubit systems.  The fringes
measured for a single driven double quantum dot (DQD)-cavity setting and
the enlarged splittings of the hybrid Floquet states in the presence of a
second DQD are well reproduced with our model. This opens a path to study
Floquet states of multi-qubit systems with arbitrarily strong coupling and
reveals a new perspective for understanding strongly driven hybrid systems.
\end{abstract}

\maketitle

Semiconductor quantum dots (QDs) coupled to superconducting cavities
provide a platform for investigating and exploiting light-matter interactions
\cite{Gu2017, Burkard2020} with a high potential for applications in
solid-state quantum information processing. Since the coupling
strength between the cavity and the qubits determines the speed of gate operations
and information exchange \cite{Blais2004, Gu2017, Burkard2020,
Blais2021}, the development of cQED settings with strong interaction is
of high interest. Experimental progress in QD-based cQED, such as
high-impedance superconducting cavities \cite{Samkharadze2016,
Stockklauser2017}, greatly increased the coupling strength, allowing
systematic investigations of the physics of the Jaynes-Cummings model
\cite{Viennot2015, Stockklauser2017, Mi2017, Samkharadze2018, Landig2018,
Mi20182, Bonsen2022}, the quantum Rabi model \cite{Scarlino2021}, and
topology \cite{PerezGonzalezPCCP22}.  Moreover, it provides a direct path to
integrate multiple qubits.

Strong periodic driving is a powerful and widely used tool in quantum
control \cite{SilveriRPP17, Ivakhnenko2023}, quantum simulation \cite{Goldman2014,
Kyriienko2018}, and system characterization \cite{Berns2008, Stehlik2012,
Forster2014, Gonzalez-Zalba2016, Bogan2018}. Investigation of the
corresponding Floquet dynamics is crucial for understanding such strongly
driven systems \cite{Ivakhnenko2023} and provides a solid foundation for
further improvements in practical applications \cite{Kyriienko2018,
Mundada2020, Rudner2020}. Recently, Floquet spectroscopy \cite{Koski2018}
and the stationary Floquet state \cite{Chen2021} of a driven DQD have been
explored via a dispersively coupled cavity. Motivated by experimental
advances, a theory for dispersive cavity readout of driven quantum systems has
been proposed \cite{Kohler2017, Kohler2018}, restricted to settings with
weak system-cavity coupling strengths within the linear-response limit.
However, the weak coupling regime, which most of the experimental and
theoretical works focused on, cannot meet the emerging need for large
coupling strengths to perform coherent quantum information exchange and
scalable quantum networks \cite{Burkard2020, Blais2021, Kimble2008}. In
addition, despite that a few works have studied two-DQD-cavity systems in the
context of the Tavis-Cummings model and cavity-mediated long-range coupling
between qubits \cite{VanWoerkom2018, Scarlino2019, Landig2019, Borjans2020,
Wang2021, Harvey-Collard2022}, the dynamics of a driven multiqubit-cavity
system remains unexplored.

In this work, we demonstrate a strongly driven hybrid system consisting of
two spatially separated GaAs DQDs coupled to a superconducting NbTiN
cavity. Benefiting from the enhanced coupling strength for the
high-impedance cavity, the system is working beyond the scope of the existing
theories \cite{Kohler2017, Kohler2018}. Different from these theories which
treat the DQD as a relatively independent strongly driven system for the
weak coupling strength, here we further develop a generalized theory by
considering the Floquet states of the full hybrid system. In doing so, we
treat the cavity as part of a central driven system, which provides an
approach applicable for arbitrarily strong DQD-cavity coupling and
also captures the cavity-mediated interaction between different DQDs. In
our experiment, Landau-Zener-St\"uckelberg-Majorana (LZSM) interference
patterns and splittings for a driven single DQD-cavity setup as well as the
enlarged splittings for two-DQD-cavity system are experimentally observed
in the cavity transmission. The results are analyzed and well reproduced
with our model.

\begin{figure}
    \includegraphics{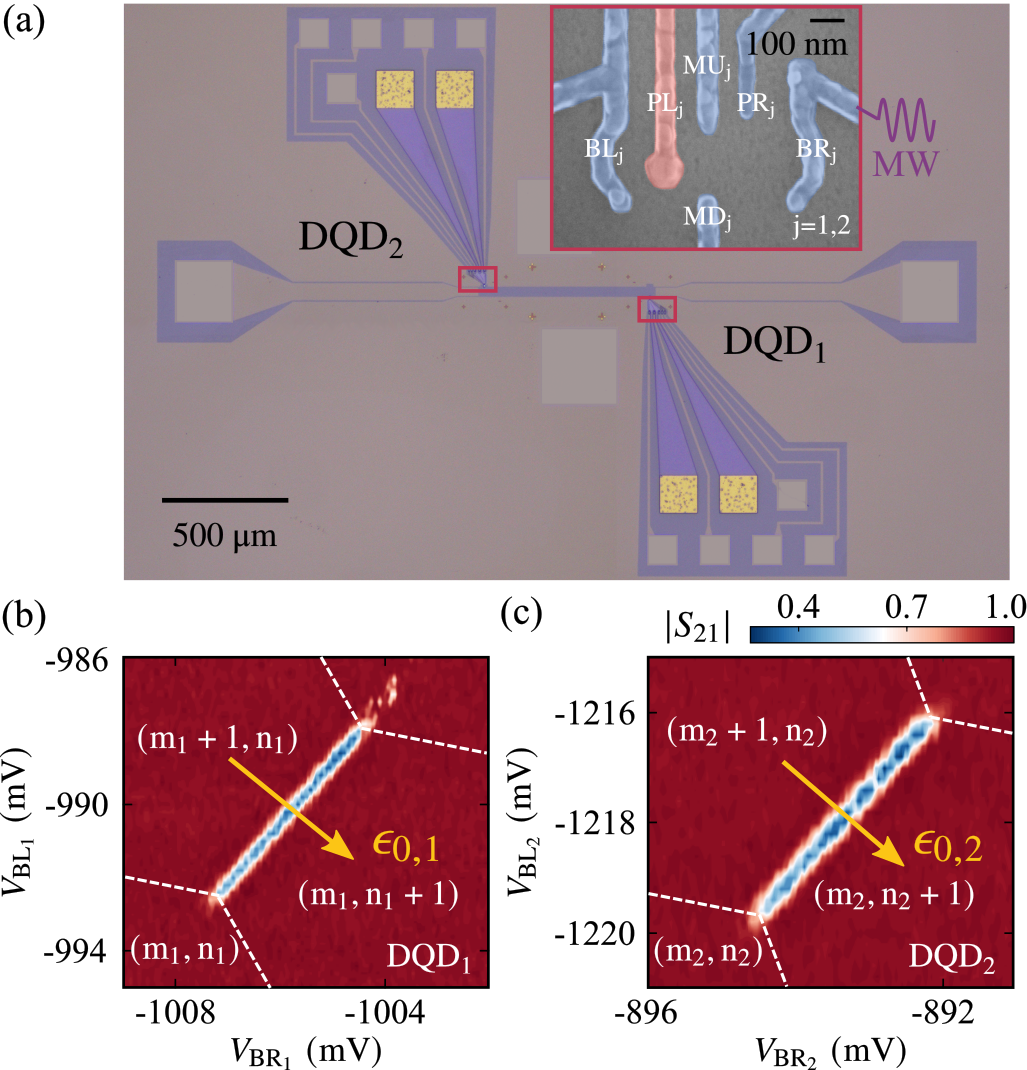}
    \caption{\label{fig:1} (a) Optical micrograph of the device which is cooled to a temperature of $\sim\SI{20}{\milli\kelvin}$. Inset: false-color scanning electron micrograph of ${\text {DQD}}_j$. The plunger gate ${\text {PL}}_j$ (red) is connected to the cavity. (b)-(c) Charge stability diagrams measured by cavity transmission amplitude $|S_{21}|$ as a function of gate voltages for ${\text {DQD}}_1$ and ${\text {DQD}}_2$, respectively.}
\end{figure}

Figure~\ref{fig:1}(a) shows the half-wavelength superconducting NbTiN
transmission cavity containing two DQDs (${\text {DQD}}_j$, $j=1,2$)
separated by a distance of roughly $\SI{670}{\micro m}$. Each
DQD is connected to either voltage antinode of the high-impedance
($Z_r\approx \SI{2}{k\Omega}$) cavity with a center frequency
$\omega_c/2\pi =\SI{5.196}{GHz}$ and photon decay rate $\kappa/2\pi =
\SI{12.0}{MHz}$.

The DQDs are formed in a GaAs/AlGaAs quantum well with gate electrodes
[inset in Fig.~\ref{fig:1}(a)]. The occupation number $(m_j, n_j)$ in
$\text{DQD}_j$ is controlled by gate voltages $V_{\text{BR}_j}$
and $V_{\text{BL}_j}$, as displayed in Figs.~\ref{fig:1}(b) and
\ref{fig:1}(c). An excess electron in $\text{DQD}_j$ forms a charge qubit
described by the Hamiltonian
\begin{equation}
    H_{q,j} = \frac{\epsilon_j}{2}\sigma_{z,j} + t_j\sigma_{x,j}.
\end{equation}
Here $\sigma$ denotes Pauli matrices, while $\epsilon_j$ is the energy
detuning between the left and right dot of ${\text {DQD}}_j$, which can be
adjusted by $V_{\text{BR}_j}$. The interdot tunnel coupling
$2t_j$ can be tuned via $V_{\text{MU}_j}$ and $V_{\text{MD}_j}$.

The hybrid system is modeled by the Hamiltonian
\begin{equation}
H(t) = \sum_j H_{q,j}(t) + \sum_j g_j Z_j (a^\dagger+a)
+ \hbar\omega_c a^\dagger a ,
\label{H}
\end{equation}
where $a$ $(a^\dagger)$ is the annihilation (creation) operator of a cavity
photon, and $H_{q,j}(t)$ refers to ${\text {DQD}}_j$.
Its scaled dipole operator $Z_j = \sigma_{z,j}$ couples to the electric
field of the cavity with strength $g_j$. The coupling strength between
${\text {DQD}}_1$ and the cavity is estimated to reach $g_{1}/2\pi =
\SI{85}{MHz}$ at $2t_{1}/h\approx\SI{5.2}{GHz}$ with decoherence rate
$\gamma_1/2\pi \approx \SI{90}{MHz}$, while $g_{2}/2\pi = \SI{80}{MHz}$ at
$2t_{2}/h\approx\SI{5.16}{GHz}$ with $\gamma_2/2\pi \approx\SI{100}{MHz}$
for ${\text {DQD}}_2$.
Experimentally, the continuous microwave is applied to gate ${\text {BR}}_j$
to periodically drive the system such that
\begin{equation}
    \epsilon_j(t)=\epsilon_{0,j}+A_{d,j} {\sin}(2\pi f_{d}t)
    \label{epsit}
\end{equation}
with offset $\epsilon_{0,j}$, driving amplitude $A_{d,j}$, and driving
frequency $f_d \equiv \Omega/2\pi$. We study the dynamics of the driven
system by probing the transmission signal $|S_{21}|$ through the cavity.

To establish a theory for the transmission of a cavity coupled to various
driven DQDs, one may extend the approach of Refs.~\cite{Kohler2017,
Kohler2018, Mi2018, Chen2021} and consider the action of the cavity on each
DQD and its backaction with non-equilibrium linear response theory.  Since
this approach is based on second order perturbation theory in the weak
DQD-cavity couplings, no cross terms between different DQDs occur, such
that one can compute the impact of each DQD separately.  Therefore, the
shift of the cavity resonance, which governs the transmission, simply
follows by summing the contributions of the individual DQDs.  

\begin{figure}
    \includegraphics{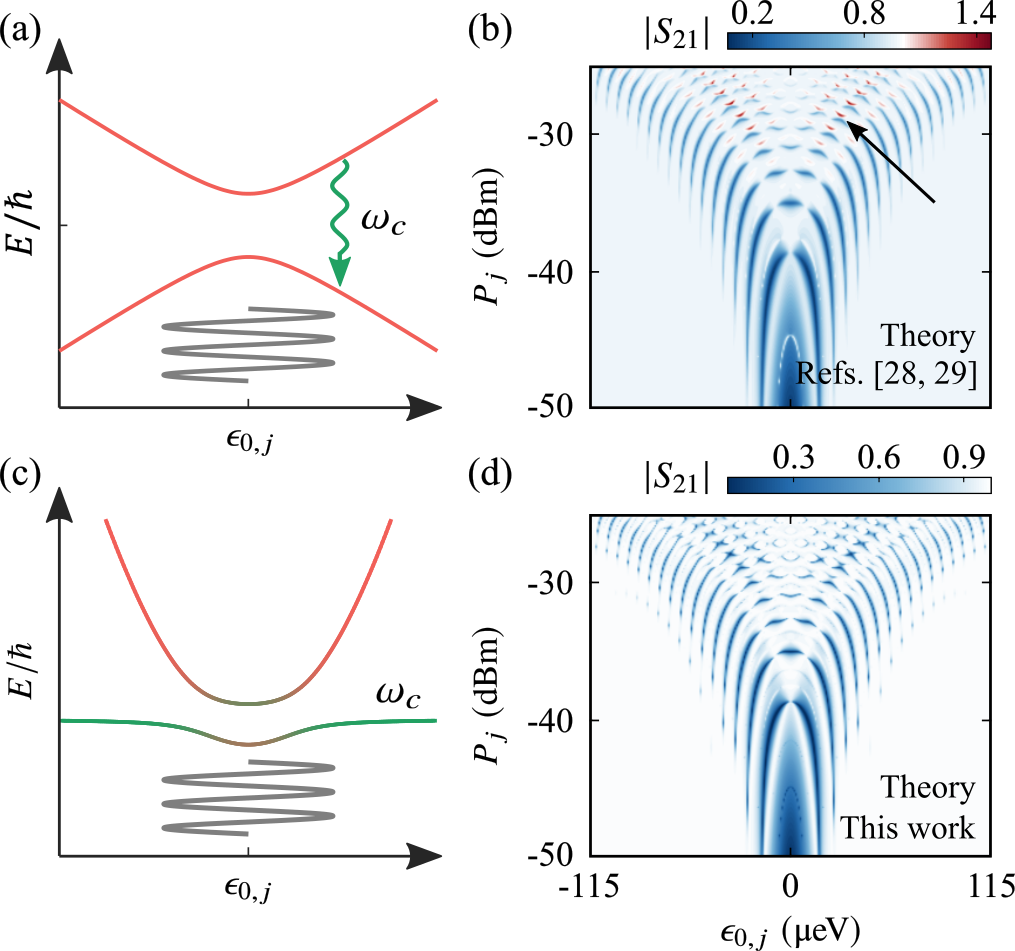}
    \caption{\label{fig:2} The energy level schematic diagram of (a) a single DQD and (c) single-DQD-cavity system. The gray oscillating curves represent the microwave drivings. The red (green) color denotes the DQD (cavity) state, and the line color reflects the state hybridization due to the coupling. 
    The calculated results for $|S_{21}|$ at $\omega_p=\omega_c$ obtained with theory (b) in the previous work and (d) in this work. $P_j = \SI{-40}{dBm}$ corresponds to $A_{d,j}=\SI{20.2}{\micro eV}$ in the simulation. The other simulation parameters are $2t_j = \SI{5.2}{GHz}$, $g_j/2\pi= \SI{85}{MHz}$, $f_d=\SI{1.4}{GHz}$.}
\end{figure}

Figure \ref{fig:2}(a) visualizes this viewpoint for a single DQD. Within each
driving period, the two relevant DQD states pick up a relative phase
determined by the difference of the Floquet quasienergies.  When it matches a
multiple of $2\pi$, one observes fringes in the excitation probability
\cite{Ivakhnenko2023}.  The resonance condition for the cavity signal
involves the cavity frequency and reads $\Delta\mu = \omega_c + k\Omega$ 
with integer $k$ \cite{Kohler2018}.  For a detailed 
discussion of these competing resonance conditions and their experimental 
verification, see Ref.~\cite{Chen2021}.

The computed interference pattern for the parameters of our setup is shown
in Fig.~\ref{fig:2}(b).  Its fringes comply with the resonance condition,
but for some values of the driving parameters the theory predicts
transmissions considerably larger than unity (red areas, marked by an
arrow), which may indicate lasing \cite{MarthalerPRB15, LiuS15,
NeilingerPRB16}.  Here, however, this is not the case.  It is rather such
that the DQD-cavity coupling strength is beyond the linear response limit.
Moreover, cavity-mediated interactions between the DQDs are ignored. To
overcome these shortcomings, we develop a theory for the readout of driven
qubits in which the cavity is considered as part of the central system.
Figure \ref{fig:2}(c) illustrates this idea for a single DQD coupled to a
cavity. This will allow us to treat settings with arbitrarily strong
DQD-cavity coupling.  For details of the derivation, see the \supplement.

We start from the quantum Langevin equation for the cavity field $a$ with
an inhomogeneity that corresponds to the Hamiltonian
\begin{equation}
H_1(t) = -i\sum_{\nu=1,2} \sqrt{\kappa_\nu} a^\dagger
a_{\text{in},\nu}(t) + \text{H.c.}
\label{Hin}
\end{equation}
with the incoming fields $a_{\text{in},\nu}$ at port $\nu=1,2$.  The
corresponding time-reversed equation relates incoming and outgoing fields
as $a_{\text{out},\nu} - a_{\text{in},\nu} = \sqrt{\kappa_\nu} a$
\cite{CollettPRA84, Blais2004}.  Thus, to obtain the transmission, we have
to compute how the $a_{\text{in},\nu}$ affects the cavity operator $a$.

To this end, we employ non-equilibrium linear response theory for the
perturbation caused by $H_1$.  Since the cavity is probed at or close to
resonance, $\omega_p \approx\omega_c$, the Hermitian conjugate contribution
is off-resonant and, thus, can be neglected, such that the inputs act only
via the cavity operator $a^\dagger$.  In agreement with Kubo formula, we
find that the perturbation $a_{\text{in},\nu}(t)$ and the response $\langle
a(t)\rangle$ are linked by the susceptibility
\begin{equation}
    \chi(t,t') = -i\langle [a(t),a^\dagger(t')]\rangle_0\theta(t-t'),
    \label{chinew}
\end{equation}
with the Heaviside step function $\theta$.  Notice that here the
expectation value $\langle\cdots\rangle_0$ considers the driven dissipative
dynamics of the full DQDs-cavity compound.  Since the DQDs are driven, the
response depends explicitly on both times.  Then $\langle a(t)\rangle$ is
no longer given by a simple convolution, but acquires a summation over
Fourier components $\chi^{(k)}(\omega)$. Nevertheless, for the
experimentally relevant time-averaged cavity signal, knowledge of the
component with $k=0$ is sufficient \cite{Kohler2017, Kohler2018}.
Taking this average, the response in
frequency space reads $\langle a_{\omega}\rangle = -i\sum_\nu
\sqrt{\kappa_\nu} \chi^{(0)}(\omega) a_{\text{in},\nu}(\omega)$.  For an
input at port $\nu=1$ only, the input-output relation directly provides the
time-averaged cavity transmission
\begin{equation}
    S_{21}(\omega) = -i\sqrt{\kappa_1\kappa_2}\chi^{(0)}(\omega)
    \label{s21}
\end{equation}
and the reflection $S_{11}(\omega) = 1 - i\kappa_1\chi^{(0)}(\omega)$.

The remaining task is the computation of $\chi^{(0)}(\omega)$ for which we
proceed as in Refs.~\cite{Kohler2018, Chen2021}, but with the DQD
Hamiltonian replaced by the full DQDs-cavity Hamiltonian \eqref{H}.
Following this scheme, we first compute the Floquet states
$|\phi_\alpha(t)\rangle$ of $H(t)$ and the quasienergies $\mu_\alpha$ by
solving the eigenvalue equation $[H(t)-i\hbar d/dt]|\phi(t)\rangle =
\mu|\phi(t)\rangle$.  The result is then used to evaluate the
dissipative kernel of the Bloch-Redfield equation, which yields the
transition rates between Floquet states and, hence, the steady-state
populations of the Floquet states, $p_\alpha$. With these ingredients, the
susceptibility in Eq.~\eqref{s21} becomes
\begin{equation}
    \chi^{(0)}(\omega)=\sum_{\alpha,\beta,k}
    \frac{(p_\alpha-p_\beta) |a_{\alpha\beta,k}|^2}
         {\omega +(\mu_\alpha-\mu_\beta)/\hbar - k\Omega+i\kappa/2},
    \label{chi0}
\end{equation}
where $a_{\alpha\beta,k}$ is the $k$th Fourier component of the transition
matrix element $a(t)=\langle \phi_\alpha(t)|a|\phi_\beta(t)\rangle$. The
total cavity decay rate $\kappa=\kappa_1+\kappa_2+\kappa_{\text {int}}$
consists of contributions from each port $\kappa_\nu$ and the internal
losses $\kappa_{\text {int}}$.

\begin{figure}
    \includegraphics{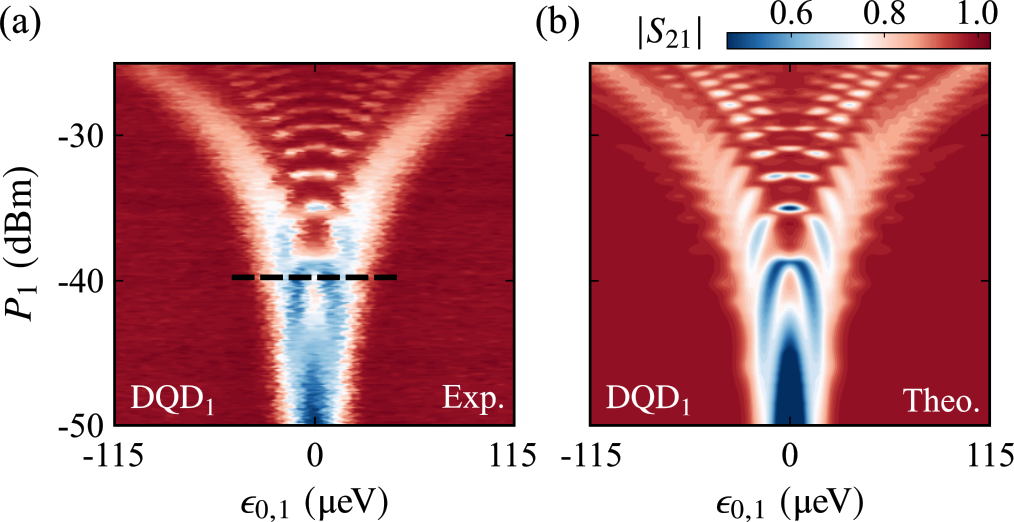}
    \caption{\label{fig:3}(a) Measured transmission $|S_{21}|$ as a function of detuning $\epsilon_{0, 1}$ and driving power $P_{1}$ for ${\text {DQD}}_1$ at $2t_1/h\approx \SI{5.2}{GHz}$ with $f_{d}= \SI{1.4}{GHz}$. (b) The corresponding theoretical result. To consider the fluctuations of $\epsilon_1$ and $P_{1}$, we convolve the plot with a Gaussian distribution of width $\sigma_{\epsilon}= \SI{4}{\micro eV}$ along the $\epsilon_{0,1}$ axis and $\sigma_{P}= \SI{0.1}{dB}$ along the $P_{1}$ axis \cite{Forster2014, Stehlik2016,Chen2021}.}
\end{figure}

The corresponding theory prediction for a single DQD is shown in
Fig.~\ref{fig:2}(d).  In contrast to the theory of Refs.~\cite{Kohler2018,
Chen2021} [Fig.~\ref{fig:2}(b)], it obeys $|S_{21}|\leq 1$ in the whole
range considered, which underlines the applicability of our theory for
values of $g_j$ beyond the linear-response regime.

To demonstrate that the preceding theory enables a quantitative understanding
of measurement results, firstly, ${\text {DQD}}_1$ is driven and coupled to
the cavity, while ${\text {DQD}}_2$ is far detuned, $\epsilon_{0,2}\gg
\hbar\omega_c$, and hence is inactive. Figure \ref{fig:3}(a) shows the
measured $|S_{21}|$ as a function of detuning $\epsilon_{0,1}$ and driving
power $P_{1}\propto A_{d,1}^2$, which maps out a LZSM interference pattern.
Within the $|\epsilon_{0,1}|<A_{d,1}$ region, a series of interference
fringes with amplitude minima significantly below unity are observed. When
the states of the ${\text {DQD}}_1$-cavity system
interfere constructively, the excited Floquet state is effectively
populated, resulting in a minima value of the numerator in Eq.~\eqref{chi0}
and a reduction in $|S_{21}|$. Figure \ref{fig:3}(b) displays the
theoretical result after convolution with a Gaussian that captures the
inhomogeneous broadening \cite{Forster2014, Stehlik2016, Chen2021}.  It is
in very good agreement with the experimental data.  Likewise, an equivalent
experiment is carried out for the case of ${\text {DQD}}_2$ coupled with
the cavity. The result is consistent with Fig.~\ref{fig:3} and is shown
in the \supplement.

\begin{figure}
    \includegraphics{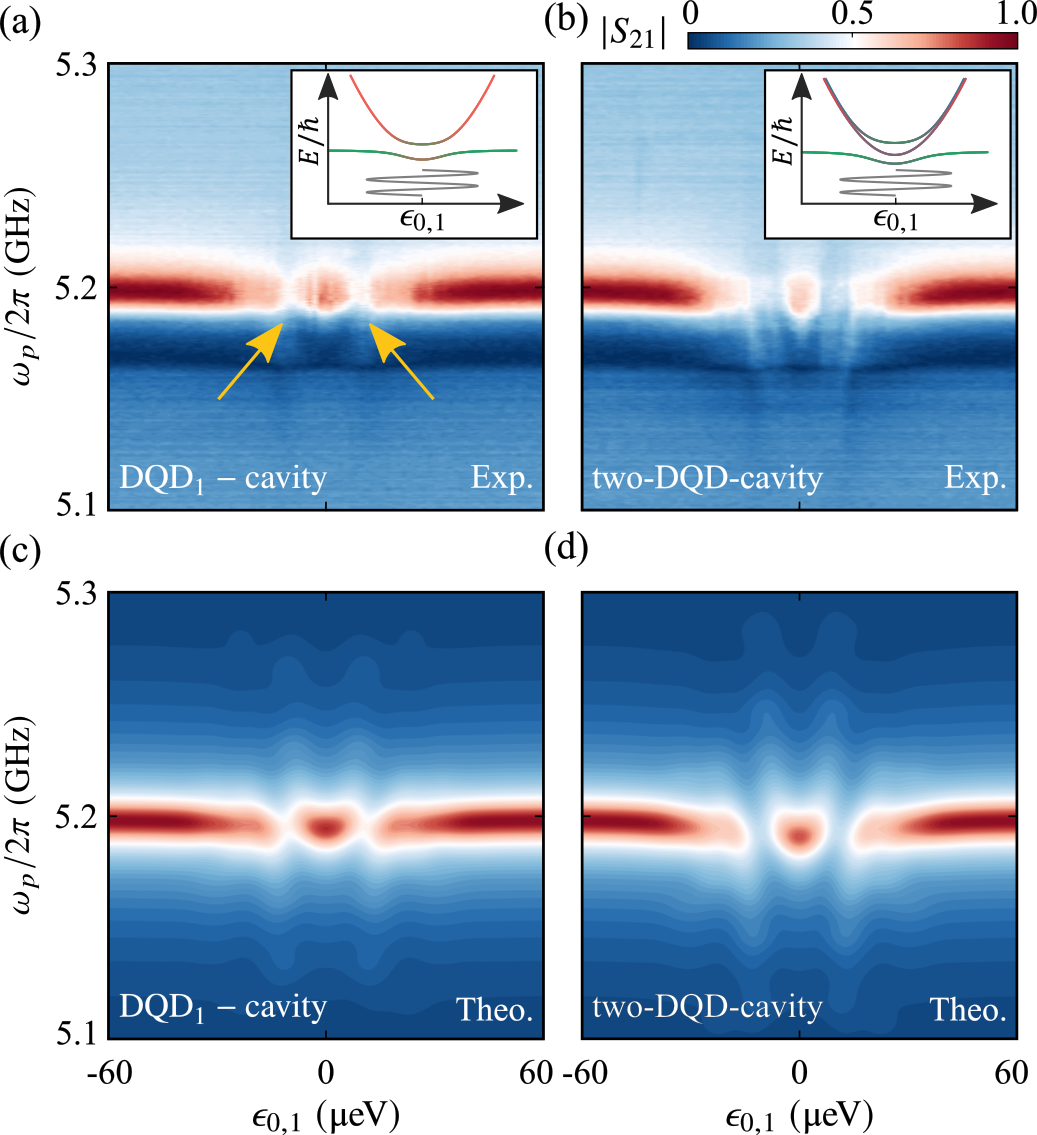}
    \caption{\label{fig:4} The $|S_{21}|$ as a function of probe frequency $\omega_p/2\pi$ and detunings $\epsilon_{0,1}$ for the (a) ${\text {DQD}}_1$-cavity system and (b) two-DQD-cavity system. In (b), $\epsilon_{0,2}$ is simultaneously varied in the range of $(-68.3,68.3)\si{\micro eV}$. $P_1=\SI{-39.9}{dBm}$ and $P_2=\SI{-42}{dBm}$ are fixed.  Insets: the energy level schematic diagram of the hybrid system. The red, blue and green colors denote the $\text{DQD}_1$, $\text{DQD}_2$ and cavity states, respectively. (c)-(d) The theoretical results after convolution with the same parameters in Fig.~\ref{fig:3}.}
\end{figure}

Scalable quantum information processing requires multiple qubits that
interact with a cavity. As a reference,
Fig.~\ref{fig:4}(a) shows the resulting measured $|S_{21}|$ for the ${\text
{DQD}}_1$-cavity system as a function of the cavity probe frequency
$\omega_p/2\pi$ and the detuning $\epsilon_{0,1}$. The drive power is fixed
at $P_{d,1}=\SI{-39.9}{dBm}$, which is marked by a black dashed line in
Fig.~\ref{fig:3}(a).  The red bar with $|S_{12}|\approx 1$ is observed when
the cavity is probed at resonances, $\omega_p=\omega_c$.  In the spectral
picture, this corresponds to an excitation energy $\hbar\omega_c$ sketched
by the green line in the inset of Fig.~\ref{fig:4}(a).  When the cavity
frequency comes close to resonance with the DQD, the cavity and the DQD are
hybridized, illustrated by the line color in the inset of Fig.~\ref{fig:4}(a). 
Then the driving leads to interference which induces a redistribution of
DQD-cavity Floquet states, which is visible as gaps in the red bar with
maximal transmission around $\epsilon_{0,1}=0$ marked by yellow arrows. The
corresponding data for ${\text {DQD}}_2$ is shown in Fig.~S1 of the
\supplement.

We now turn to the case of a driven two-DQD-cavity hybrid system by tuning both
DQDs close to resonance with the cavity and applying to both microwaves
with the same frequency $f_d=\SI{1.4}{GHz}$ at gates $\text{BR}_1$ and
$\text{BR}_2$. By simultaneously changing the detunings $\epsilon_{0,1}$
and $\epsilon_{0,2}$, we measure the transmission $|S_{21}|$ depicted in
Fig.~\ref{fig:4}(b).  Notably in comparison with the single DQD case
[Fig.~\ref{fig:4}(a)], the gaps in the red bar become significantly larger.

In contrast to the single-DQD-cavity system, the two-DQD-cavity system is a
multilevel system whose eigenenergies are schematically illustrated in the
inset of Fig.~\ref{fig:4}(b). Nevertheless, the Floquet states acquire
different phases during the driving period and interfere similarly,
resulting in a change in the population distribution and a more pronounced impact
on $|S_{21}|$.  As is indicated by Eq.~\eqref{chi0}, besides the
population, $|a_{\alpha\beta,k}|^2$ and $\omega_p
+(\mu_\alpha-\mu_\beta)/\hbar - k \Omega$ also play a role in the signal.
The increased hybridization of cavity photon with two DQDs leads to a
decrease in $|a_{\alpha\beta,k}|^2$ and a larger deviation of the energy
splitting $(\mu_\beta-\mu_\alpha)/\hbar + k \Omega$ from $\omega_p$.
Therefore, enlarged splittings are observed. All the experimental features
are well reproduced by the theoretical results in Figs.~\ref{fig:4}(b) and
\ref{fig:4}(d), which underlines that our approach is general and scalable. 

In conclusion, we have investigated both experimentally and theoretically
the driven dynamics of hybrid system in which two DQDs are strongly coupled
to a cavity.  For the theoretical description, we have developed a method
for the cavity transmission in which the cavity is considered as part of a
Floquet system. This extends the method of Ref.~\cite{Kohler2017} to cases
with DQD-cavity coupling beyond the linear-response limit. Moreover, it
allows one to treat multiple qubits that interact via the cavity. On a
quantitative level, we have demonstrated the excellent agreement of
computed and measured LZSM patterns in a regime in which limitations of the
former approach become visible.
Our approach is applicable to cQED architecture built of other physical
systems, such as semiconductor QDs in other host materials and
superconducting qubits. Our results provide a more profound insight into
the dynamics of Floquet states and may motivate future applications in
scalable hybrid quantum systems.

\begin{acknowledgments}
This work was supported by the National Natural Science Foundation of China
(Grant Nos.\ 61922074, 92265113, 12074368, and 12034018), by the Spanish
Ministry of Science, Innovation, and Universities (Grant No.\
PID2020-117787GB-I00), and by the CSIC Research Platform on Quantum Technologies
PTI-001.  This work was partially carried out at the USTC Center for Micro- and Nanoscale Research and
Fabrication.
\end{acknowledgments}

\appendix
\section{Complementary data}
\label{individualDQD2}

\begin{figure}
    \includegraphics{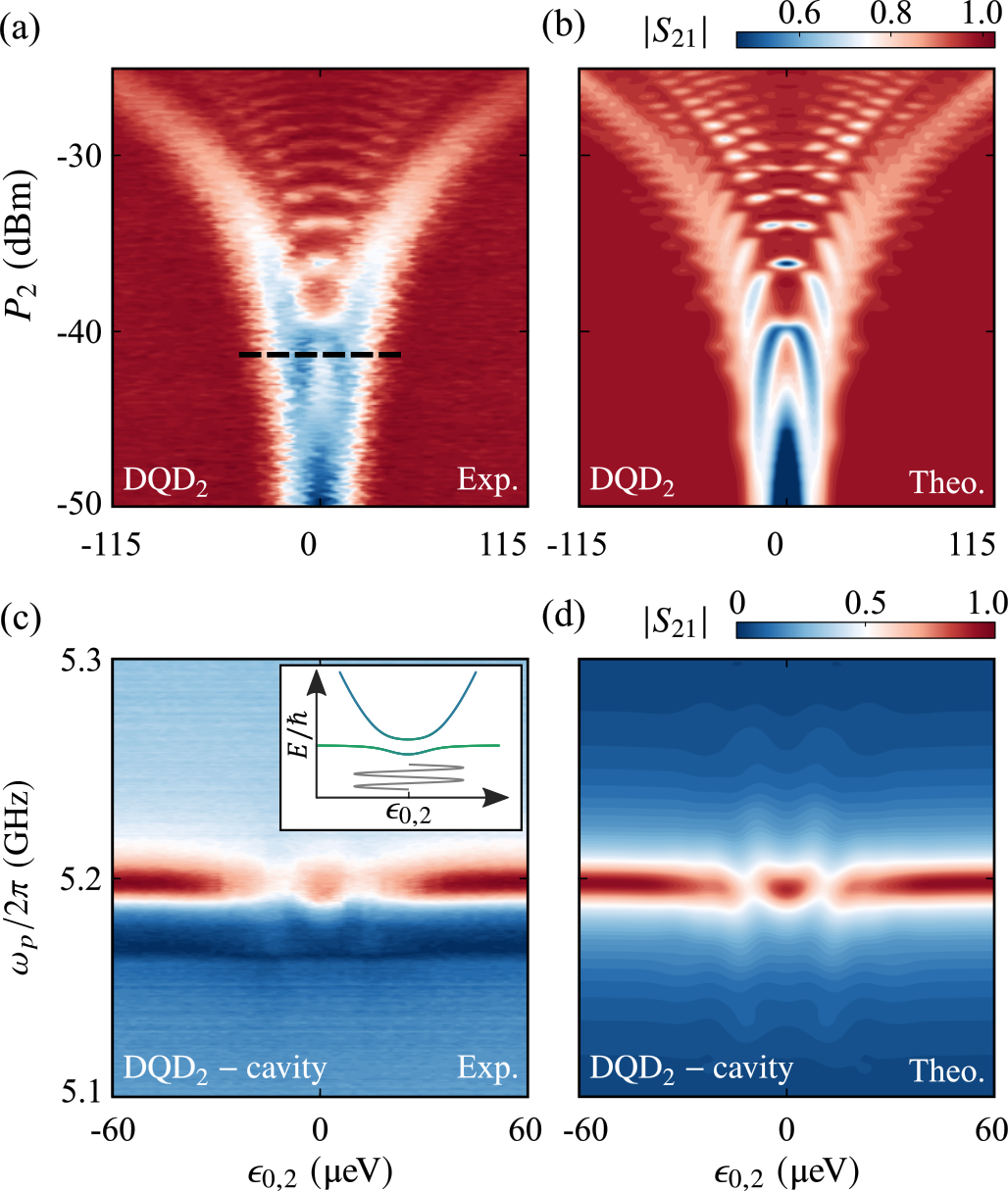}
    \caption{\label{fig:6} (a) Transmission amplitude $|S_{21}|$ as a function
of detuning $\epsilon_{0, 2}$ and drive power $P_{2}$ for ${\text {DQD}}_2$ at
tunnel coupling $2t_2/h\approx \SI{5.16}{GHz}$ with drive frequency $f_{d}=
\SI{1.4}{GHz}$. (b) Theoretical result, where the microwave power
$P_{2} = \SI{-40}{dBm}$ corresponds to the driving amplitude
$A_{d,2}=\SI{23.1}{\micro eV}$.
(c) The $|S_{21}|$ as a function of probe frequency $\omega_p/2\pi$ and
detunings $\epsilon_{0,2}$ for the driving power $P_{2}=\SI{-42}{dBm}$ marked by the black dashed line in (b). Inset:
schematic energy level diagram of the ${\rm DQD}_2$-cavity system. 
The blue (green) color denotes the DQD (cavity) state, 
and the line color reflects the mixing degree due to the coupling.
(d) Corresponding theoretical result. All theory data consider
inhomogeneous broadenings by a convolution with a Gaussian of widths
$\sigma_{\epsilon}= \SI{4}{\micro eV}$ and $\sigma_{P}=
\SI{0.1}{dB}$.}
\end{figure}

In addition to the data shown in Figs.~3 and 4 of the main text, we
present the result of the equivalent experiment carried out for the ${\rm
DQD}_2$-cavity system. We apply a continuous microwave at frequency
$f_d=\SI{1.4}{GHz}$ to ${\rm BR}_2$ gate with ${\rm DQD}_1$ far detuned.
Figure~\ref{fig:6}(a) depicts the LZSM interference of the ${\rm DQD}_2$-cavity
system. Figure~\ref{fig:6}(c) shows the transmission as a function of
$\epsilon_{0,2}$ and $\omega_p$ for the driving power fixed at the value
marked by a dashed black line in Fig.~\ref{fig:6}(a).
Gaps are observed at $\epsilon_{0,2} \approx \SI{-12.4}{\micro eV}$ and
$\SI{12.4}{\micro eV}$. The measured results
for the driven ${\rm DQD}_2$-cavity system are theoretically well
reproduced [Figs.~\ref{fig:5}(b,d)].  They look quite similar to those for
the driven ${\rm DQD}_1$-cavity system.

\section{Readout theory with cavity as part of system}
\label{appendixtheory}

We derive the transmission and the reflection of a cavity strongly coupled
to one or several periodically driven DQDs.  In Ref.~\cite{Kohler2018},
this has been achieved by a perturbation theory in the cavity-DQD couplings
which naturally holds only for sufficiently weak coupling.  To overcome
this limitation, we here treat the cavity-DQDs setting as an entire central
system and employ a perturbation theory for the usually weak cavity input.

\subsection{The cavity-DQD model}

Our starting point is the quantum Langevin equation for a cavity mode $a$
with frequency $\omega_c/2\pi$ with ports
$\nu=1,2$ coupled to one or several DQDs labeled by $j$ \cite{Blais2004},
\begin{equation}
    \dot{a}= -i\omega_c a-i\sum_j g_jZ_j-\frac{\kappa}{2}a
    -\sum_{\nu=1,2}\sqrt{\kappa_\nu}a_{{\rm{in}},\nu}(t),
    \label{a}
\end{equation}
where $Z_j$ denotes the DQD observables (usually the dipole operators)
through which the DQD is coupled with strength $g_j$ to the cavity mode.
The total cavity loss rate $\kappa=\kappa_1+\kappa_2+\kappa_{\rm int}$
consists of a $\kappa_\nu$ for each port and an internal loss rate
$\kappa_\text{int}$.  Formally, the latter may be considered as stemming
from a third port, where the corresponding input is a vacuum field which at
our level of approximation does not contribute to $\langle a\rangle$.  The
corresponding time-reversed equation, provides the input-output relation
\cite{CollettPRA84}
\begin{equation}
a_{\text{out},\nu}=a_{\text{in},\nu}+\sqrt{\kappa_\nu}a.
\label{io}
\end{equation}

Our objective is to compute with linear response theory how a cavity input
with frequency $\omega \approx\omega_c$ affects the expectation value
$\langle a(t)\rangle$. Then Eq.~\eqref{io} readily provides the
outputs and, thus, the cavity transmission and reflection.

\subsection{Nonequilibrium perturbation theory}

The last term in the quantum Langevin equation \eqref{a} corresponds to
the Hamiltonian
\begin{equation}
    \begin{aligned}
        H_1(t)&=-i\hbar\sum_{v}\sqrt{\kappa_v}a^\dagger a_{{\rm in},v}+ \rm{H.c.}\\
        &\equiv \hbar\xi(t)a^\dagger+\hbar\xi^*(t)a
    \end{aligned}
\end{equation}
with $\xi(t)$ subsuming all input fields.  In the experiment, the input is
monochromatic with frequency $\omega\approx\omega_c$ and amplitude $A_0$.

The unperturbed situation is described by a time-dependent non-equilibrium
density matrix $\rho_0(t)$ which captures all other influences such as the
driving acting upon the DQDs, the cavity-DQDs coupling, as well as DQD and
cavity dissipation.  In the corresponding interaction picture, the
Liouville-von Neumann equation reads $\dot{\tilde\rho} = -(i/\hbar)[\tilde
H_1,\tilde\rho]$.  To the lowest order in $\xi$, it can be 
written in the integrated form
\begin{equation}
    \tilde{\rho}(t)=\tilde{\rho}_0(t)
    -\int_{-\infty}^{t}dt'[\tilde{a}^\dagger \xi(t')+\tilde{a}\xi^*(t'),\tilde{\rho}_0(t')].
    \label{rho}
\end{equation}
For resonant cavity driving, $\xi \sim e^{-i\omega_ct}$, the term with
$\xi^*$ is necessarily off-resonant and can be neglected unless the quality
factor of the cavity, $Q=\omega_c/\kappa$, is extremely low. Then
transforming back to the Sch\"odinger picture, we obtain the expectation value
\begin{equation}
    \langle a(t)\rangle=\int_{-\infty}^{\infty} dt'\chi(t,t')\xi(t')
    \label{atime}
\end{equation}
with the response function 
\begin{equation}
    \begin{aligned}
    \chi(t,t')
    ={}& -i \tr\left\{a\;\mathcal{U}(t,t')[a^\dagger,\rho_0(t')]\right\}\theta(t-t')\\
    ={}& -i\langle [a(t),a^\dagger(t')]\rangle_0\theta(t-t') ,
    \label{chi1}
    \end{aligned}
\end{equation}
where $\mathcal{U}$ is the propagator of the density operator in the
absence of $H_1$ and $\theta$ is the Heaviside step function.

\subsection{Periodically driven DQDs}

When the drivings of all DQDs have equal periodicity $T=1/f_{d} \equiv
2\pi/\Omega$, the susceptibility in the steady state obeys the same time
periodicity, namely $\chi(t,t') = \chi(t+T,t'+T)$.  Therefore its spectral
decomposition reads \cite{Kohler2018}
\begin{equation}
    \chi(t,t')=\sum_k\int \frac{d\omega}{2\pi} e^{-ik\Omega t-i\omega\tau}\chi^{(k)}(\omega) .
    \label{chi2}
\end{equation}
Hence the Fourier transformed of Eq.~\eqref{atime} becomes
\begin{equation}
     a(\omega)=\sum_k \chi^{(k)}(\omega-k\Omega)\xi(\omega-k\Omega).
    \label{afrequency}
\end{equation}
Making again use of the good-cavity limit and assuming that the probe
frequency is sufficiently distant from all multiples of the drive frequency
$\Omega$, one finds that only the Fourier coefficient with $k=0$ provides a
significant contribution (see also the corresponding discussion in
Ref.~\cite{Kohler2018}).  Thus,
\begin{equation}
    a(\omega)=\chi^{(0)}(\omega)\xi(\omega),
    \label{aomega}
\end{equation}
such that for monochromatic cavity input at port~1, we obtain the cavity
transmission and reflection
\begin{align}
    S_{21}(\omega)
    ={}& \frac{a_{{\rm out},2}(\omega)}{a_{{\rm in},1}(\omega)}
    =-i\sqrt{\kappa_1\kappa_2}\chi^{(0)}(\omega),
\label{transmission1}
\\
    S_{11}(\omega)
    ={}& \frac{a_{{\rm out},1}(\omega)}{a_{{\rm in},1}(\omega)}
    =1-i\kappa_1\chi^{(0)}(\omega).
\label{reflection1}
\end{align}

So far, we have used linear response theory for the perturbation by the
cavity input and used the time periodicity of the susceptibility in the
stationary limit. The remaining issue is the computation of $\chi^{(0)}$.
This task may be performed at various levels of approximation, depending on
the relevance of the different ingredients of the unperturbed dynamics.
Here we adapt the method used in Refs.~\cite{Kohler2018, Chen2021}
according to our needs.  Its main idea is to capture the coherent dynamics
of the cavity-DQDs setup by Floquet theory, while its dissipative dynamics
is treated with a Bloch-Redfield equation in rotating-wave approximation
\cite{Redfield1957, Blum1996}.

\subsection{Floquet theory}

Floquet theorem tells us that a Schr\"odinger equation with a
$T$-periodic Hamiltonian has a complete set of solutions of the form
$|\psi_\alpha(t)\rangle = e^{-i\mu_\alpha t}|\phi_\alpha(t)\rangle$, where
the Floquet states $|\phi_\alpha(t)\rangle = \sum_k e^{-ik\Omega
t}|\phi_{\alpha,k}\rangle$ obey the periodicity of the driving, while the
phase factors are determined by the quasienergies $\mu_\alpha$.  Both can
be computed from the eigenvalue equation $[H(t)-i\hbar d/dt]
|\phi(t)\rangle = \mu|\phi(t)\rangle$.  Its solution provides the coherent
propagator of the driven system, $U(t,t') = \sum_\alpha
|\psi_\alpha(t)\rangle \langle\psi_\alpha(t')|$.

Assuming that in the steady state, each Floquet state is populated with
probability $p_\alpha$ while coherences are negligible, the density
operator of the system reads $\rho_0(t) = \sum_\alpha p_\alpha
|\phi_\alpha(t)\rangle\langle\phi_\alpha(t)|$.  Then the propagator of the
density matrix, $\mathcal{U}(t,t')$, can be expressed in terms of $U(t,t')$, which
allows the direct evaluation of the susceptibility \eqref{chi1}.  Fourier
transformation with respect to $\tau=t-t'$ and time-averaging yields the
component relevant for the cavity response,
\begin{equation}
    \chi^{(0)}(\omega)
    =\sum_{\alpha,\beta,k}\frac{(p_\alpha-p_\beta)|a_{\alpha\beta}^{(k)}|^2}
    {\omega +\mu_\alpha-\mu_\beta - k\Omega +i\kappa/2},
\end{equation}
where $a_{\alpha\beta}^{(k)}$ is the $k$th Fourier component of the
$T$-periodic transition matrix element
$\langle\phi_\alpha(t)|a|\phi_\beta(t)\rangle$.  The linewidth $\kappa/2$
in the denominator follows from the Langevin equation \eqref{a} in the
absence of the DQDs. It has been introduced phenomenologically under the
assumption that the DQDs do not significantly alter the cavity dissipation.

\subsection{Bloch-Redfield equation}

To obtain a well-defined solution for the Floquet state populations, we
have to include dissipative processes.  Here we consider dissipation
stemming from the coupling of the DQDs to environmental degrees of freedom
and the cavity decay manifested in the Langevin equation \eqref{a}.

We model DQD dissipation by introducing a system-bath coupling of the
form $H_\text{tot} = H(t) + H_\text{env} + V$, where the Hamiltonians
\begin{align}
H_\text{env} ={}& \sum_q \hbar\omega_q b_q^\dagger b_q
\\
V ={}& X \sum_q \lambda_q (b_q^\dagger+b_q)
\end{align}
describe the environment and its coupling to the DQD with strengths
$\lambda_q$, respectively.  As in Ref.~\cite{Chen2021}, we couple the bath
via the tunnel operator, $X=\sigma_x$.  The impact of each bath can
be summarized in the spectral density $J(\omega) = \pi\sum_q
|\lambda_q|^2\delta(\omega_q-\omega) \equiv
\pi\alpha\omega/2$ which we assume Ohmic with dimensionless dissipation
strength $\alpha$ \cite{LeggettRMP87,HanggiRMP90, Weiss1998}.
In the case of several DQDs, each DQD is coupled to a separate bath, while
all baths are assumed to have the same spectral density.

From the Liouville-von Neumann equation $\dot\rho_\text{tot}
= -(i/\hbar) [H,\rho_\text{tot}]$ follows by second-order perturbation
theory in $V$, the Bloch-Redfield equation for the central system
\cite{Redfield1957, Blum1996}
\begin{equation}
\begin{split}
\dot\rho =& -\frac{i}{\hbar}[H(t),\rho]
\\
& - \frac{1}{\hbar^2}\int_0^\infty d\tau \tr_\text{env}
  [V,[V(t,t-\tau),\rho\otimes\rho_\text{env}]] ,
\end{split}
\label{BR}
\end{equation}
with the interaction picture operator $V(t,t') = U^\dagger(t,t') V
U(t,t')$, where the Floquet representation of the coherent propagator $U$
allows us to evaluate the time integral in Eq.~\eqref{BR}.  Within a
rotating-wave approximation, the coherent terms vanish, while the
dissipative term assumes the form
\begin{equation}
    \dot{p}_\alpha
    = \sum_\beta (w_{\alpha\leftarrow \beta}p_\beta
     -w_{\beta\leftarrow \alpha}p_\alpha).
    \label{pauli}
\end{equation}
The DQD contribution to the dissipative transition rates follows from
Eq.~\eqref{BR}, which after some algebra becomes
\begin{equation}
    w_{\alpha\leftarrow \beta}^\text{DQD}
    = 2\sum_k |X_{\alpha\beta}^{(k)}|^2 N(\mu_\alpha-\mu_\beta-k\Omega) .
\end{equation}
Here, $X_{\alpha\beta}^{(k)}$ is the $k$th Fourier component of the
$T$-periodic time-dependent transition matrix element
$\langle\phi_\alpha(t)|X|\phi_\beta(t)\rangle$, while $N(\mu) =
J(\mu)(e^{\mu/k_BT} - 1)^{-1}$ with the second factor being the usual
bosonic occupation number.

For the cavity dissipation, we follow a simpler route and employ the
Lindblad dissipator $\mathcal{D}\rho= \kappa(a\rho
a^\dagger-a^\dagger a \rho/2-\rho a^\dagger a/2)$ which corresponds to the
dissipative terms in the Langevin equation \eqref{a}
\cite{GardinerZoller2004}. By decomposing this into the Floquet basis and
utilizing the rotating-wave approximation, the transition rates become
\begin{equation}
    w_{\alpha\leftarrow \beta}^{\rm cavity} = \kappa \sum_k |a_{\alpha\beta}^{(k)}|^2 
\end{equation}
with $a_{\alpha\beta}^{(k)}$ as defined above.

The total transition rate between Floquet states in the full cavity-DQDs
space is the sum of both contributions,
\begin{equation}
    w_{\alpha\leftarrow \beta}
    = w_{\alpha\leftarrow \beta}^\text{DQDs}
      +w_{\alpha\leftarrow \beta}^\text{cavity}.
    \label{wab}
\end{equation}
Solving the master equation \eqref{pauli} with the rates \eqref{wab} yields
the steady-state populations $p_\alpha$ of the Floquet states and, thus,
the steady-state density operator $\rho_0(t) = \sum_\alpha p_\alpha
|\phi_\alpha(t)\rangle\langle\phi_\alpha(t)|$.

\section{Background subtraction and Fano effect}
\begin{figure*}[t]
    \includegraphics{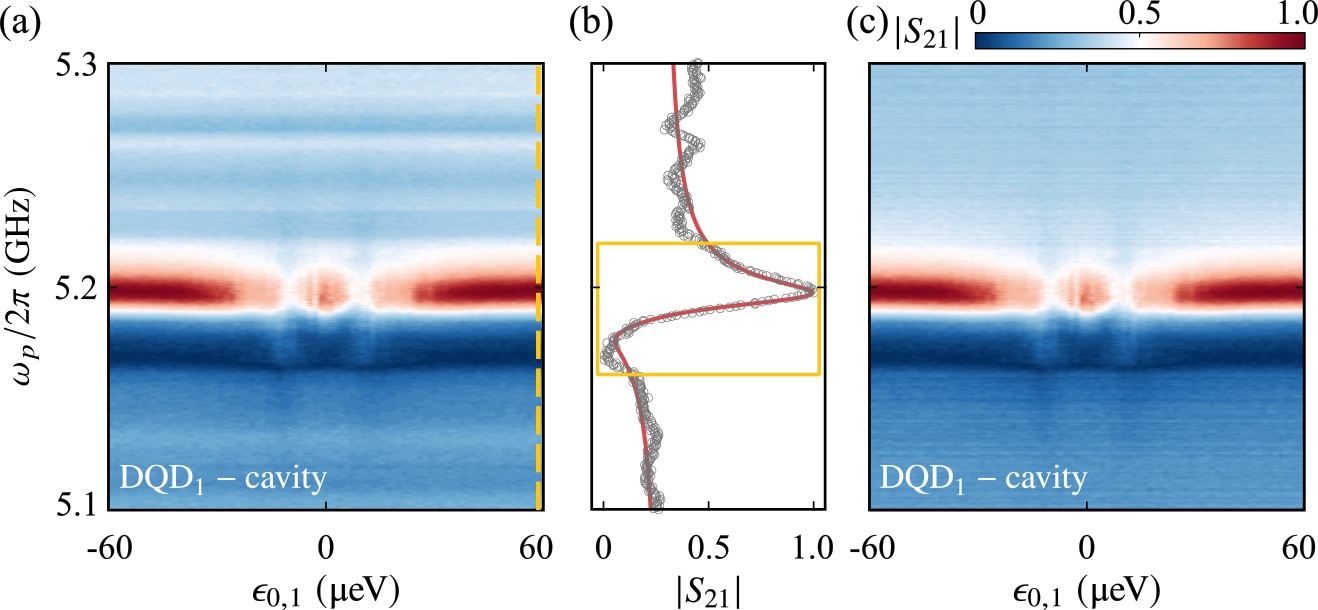}
    \caption{\label{fig:5} (a) The measured $|S_{21}|$ a function of probe frequency $\omega_p/2\pi$ and detunings $\epsilon_{0,2}$ with background.  (b) The measured $|S_{21}|$ (dots) as a function of $\omega_p/2\pi$ along the yellow dashed line in (a). The red solid line is a fit with the bare cavity transmission equation. The asymmetry of the resonance line shape is due to the Fano interference effect \cite{Leppakangas2019}. To avoid the over processing of data, we do not remove the background noise between $f_p=$\SI{5.16}{GHz} to \SI{5.22}{GHz} (the yellow rectangle). The post-processed data are plotted in (c). }
\end{figure*}
Figure~4 of the main text shows the experiment data after subtracting the
background noise. Here we show the post-processing procedure of the
experiment data. The raw data is displayed in Fig.~\ref{fig:5}(a). We find
that the background noises are identical along $\epsilon_{0,1}$ axis but
change along the $\omega_p/2\pi$ axis. We infer that this is derived from
the measurement circuit. Figure \ref{fig:5}(b) shows $|S_{21}|$ as a function
of $\omega_p/2\pi$ at $\epsilon_{0,1}\approx \SI{60}{\micro eV}$, where the
background fluctuations are more clearly seen. Since
$\epsilon_{0,1}\gg\hbar\omega_c$, the influence of ${\text {DQD}}_1$ on the
cavity is negligible, thus the data can be fitted with the equation
describing the bare cavity transmission:
\begin{equation}
    S_{21} = \frac{-i\sqrt{\kappa_1\kappa_2}}{\omega_c-\omega_p-i\kappa/2}+q.
\end{equation}
The complex constant $q$ is added to include the Fano effect due to
interference between microwave transmission through the cavity and through
a background \cite{Leppakangas2019}. We can obtain the background noise by
comparing the data and the fit, then subtract it to get the post-processed
data as shown in Figs. \ref{fig:5}(c). All the experimental data shown in
Fig.~4 of the main text and Fig. \ref{fig:6}(c) are processed in this way.

\end{document}